\documentclass[journal=jacsat, manuscript=article, notitlepage, superscriptaddress]{achemso}
\usepackage[pdftex, dvipsnames]{xcolor}
\usepackage{achemso}
\usepackage{chemformula} 
\usepackage[T1]{fontenc} 

\def\hyph{-\penalty0\hskip0pt\relax}
\usepackage{etoolbox}

\makeatletter
\renewcommand*{\acs@author@fnsymbol@symbol}[1]{
    \ifcase #1 *\or
    1\or
    2\or
    3\or
    4\or
    5\or
    \dag
    \fi
}
        
\renewcommand*\acs@contact@details{
    {\sffamily *\,E-mail: \acs@email@list }%
    \acs@number@list
}           

\patchcmd{\acs@address@list@auxii}
{\acs@author@fnsymbol{\acs@affil@marker@cnt}}
{\textsuperscript{\acs@author@fnsymbol{\acs@affil@marker@cnt}}}
{}{}

\patchcmd{\acs@address@list@auxii}
{{\acs@author@fnsymbol{\acs@affil@marker@cnt}\@nameuse{@altaffil@\@roman\@tempcnta}\par}}
{{\textsuperscript{\acs@author@fnsymbol{\acs@affil@marker@cnt}}\@nameuse{@altaffil@\@roman\@tempcnta}\par}}
{}{}     

\usepackage[T1]{fontenc} 
\usepackage{chemformula} 

\usepackage{makecell,tabularx}
\setcellgapes{3pt}
\usepackage{braket}

\usepackage[separate-uncertainty=true, 
            multi-part-units=single,
            binary-units]{siunitx} 

\DeclareSIUnit{\sample}{S}
\DeclareSIUnit\bit{b}

\usepackage{xargs}
\usepackage[disable,colorinlistoftodos,prependcaption,textsize=tiny]{todonotes}
\newcommandx{\unsure}[2][1=]{\todo[linecolor=red,backgroundcolor=red!25,bordercolor=red,#1]{#2}}
\newcommandx{\change}[2][1=]{\todo[linecolor=blue,backgroundcolor=blue!25,bordercolor=blue,#1]{#2}}
\newcommandx{\FIXME}[1][1=]{\todo[linecolor=blue,backgroundcolor=blue!25,bordercolor=blue,#1]{FIXME}}
\newcommandx{\info}[2][1=]{\todo[linecolor=OliveGreen,backgroundcolor=OliveGreen!25,bordercolor=OliveGreen,#1]{#2}}
\newcommandx{\improvement}[2][1=]{\todo[linecolor=Plum,backgroundcolor=Plum!25,bordercolor=Plum,#1]{#2}}
\newcommandx{\addref}[1][1=]{\todo[linecolor=red,backgroundcolor=red!25,bordercolor=red,#1]{Add reference}}

\newcommand\gtwo[2][]{\ifthenelse{\isempty{#2}}{$g^{(2)}(0)$}{\ifthenelse{\isempty{#1}}{$g^{(2)}(0)=#2$}{$g^{(2)}(0)=#2\pm#1$}}}

\author{Mohammed K. Alqedra}
\affiliation[1]{Quantum Nanophotonics, KTH Royal Institute of Technology, Roslagstullsbacken 21, 10691 Stockholm, Sweden}
\altaffiliation{Equal contribution}
\email{alqedra@kth.se}

\author{Chiao-Tzu Huang}
\affiliation[2]{Research Center for Critical Issues, Academia Sinica, Tainan, Taiwan}
\alsoaffiliation[5]{Department of Electrophysics, National Yang Ming Chiao Tung University, Hsinchu 30010, Taiwan}
\altaffiliation{Equal contribution}

\author{Edith Yeung}
\affiliation[3]{National Research Council of Canada, Ottawa, Ontario, Canada, K1A 0R6}
\alsoaffiliation[4]{University of Ottawa, Ottawa, Ontario, Canada, K1N 6N5}

\author{Wen-Hao Chang}
\affiliation[5]{Research Center for Critical Issues, Academia Sinica, Tainan, Taiwan}
\alsoaffiliation[2]{Department of Electrophysics, National Yang Ming Chiao Tung University, Hsinchu 30010, Taiwan}

\author{Sofiane Haffouz}
\affiliation[3]{National Research Council of Canada, Ottawa, Ontario, Canada, K1A 0R6}

\author{Philip J. Poole}
\affiliation[3]{National Research Council of Canada, Ottawa, Ontario, Canada, K1A 0R6}

\author{Dan Dalacu}
\affiliation[3]{National Research Council of Canada, Ottawa, Ontario, Canada, K1A 0R6}
\alsoaffiliation[4]{University of Ottawa, Ottawa, Ontario, Canada, K1N 6N5}

\author{Ali W. Elshaari}
\affiliation[1]{Quantum Nanophotonics, KTH Royal Institute of Technology, Roslagstullsbacken 21, 10691 Stockholm, Sweden}
\author{Val Zwiller}
\email{zwiller@kth.se}
\affiliation[1]{Quantum Nanophotonics, KTH Royal Institute of Technology, Roslagstullsbacken 21, 10691 Stockholm, Sweden}

\title{On-demand generation of entangled photons pairs in the telecom O-band from nanowire quantum dots}

\begin{document}

\newpage
\begin{abstract}
On-demand entangled photon pairs at telecom wavelengths are crucial for quantum communication, distributed quantum computing, and quantum-enhanced sensing and metrology. The O-band is particularly advantageous because of its minimal chromatic dispersion and transmission loss in optical fibers, making it well-suited for long-distance quantum networks. Site-controlled nanowire quantum dots have emerged as a promising platform for the on-demand generation of single and entangled photons, offering high extraction efficiency and the potential for scalable fabrication in large uniform arrays. However, their operation has been largely restricted to the visible and first near-infrared (NIR-I) windows. Here, we demonstrate an on-demand bright source of entangled photon pairs with high fidelity in the telecom O-band based on site-controlled nanowire quantum dots. We measure a fine-structure splitting of 4.6 $\mu$eV, verifying the suitability of the quantum dot for generating high-fidelity polarization-entangled photon pairs. Full quantum state tomography of the two-photon state generated by the biexciton\hyph exciton cascade reveals a maximum fidelity of $85.8\% \pm 1.1\%$ to the $\Phi^+$ Bell state, and a maximum concurrence of $75.1\% \pm 2.1\%$. We estimate the source efficiency at the first lens to be 12.5$\%$. This bright, scalable, and deterministic source of entangled photons in the telecom range represents a valuable step forward in advancing practical quantum applications at telecom wavelengths.                      
\end{abstract}

\maketitle

Entanglement is a key resource for various quantum technologies, including distributed quantum computing \cite{Cirac1999Jun}, quantum communication\cite{Kimble2008Jun, Wehner2018Oct} and distributed quantum sensing \cite{Guo2020Mar}. In quantum communication, entanglement enables protocols that can securely transmit information across long distances. When implemented effectively, it can extend the range of transmission by leveraging quantum repeaters and entanglement swapping techniques \cite{Briegel1998Dec, Zukowski1993Dec}. For long-distance quantum communication over optical fibers, the telecom O-band (1260–1360 nm) is especially advantageous due to its near-zero chromatic dispersion and low transmission losses in standard fiber networks, making it suitable for quantum networks that aim to achieve high fidelity over large distances. Crucially, achieving these goals requires a reliable on-demand source of entangled photons, capable of generating photon pairs deterministically and with high efficiency \cite{Neuwirth2021Dec, Vajner2022Jul, Basset2021Mar}. Such sources overcome the limitations of probabilistic emitters, ensuring synchronized operation with quantum repeaters and maximizing the scalability of quantum networks.

Photon-pair sources based on spontaneous parametric down-conversion (SPDC) have long served as the workhorse for demonstrations of fiber- and satellite-based quantum communication \cite{Kwiat1995Dec,Yin2017Jun}. However, as probabilistic sources, SPDC-based systems emit entangled photon pairs at random intervals, which limits their scalability and suitability for certain applications. Specifically, the random nature of SPDC emission complicates synchronization and hinders efficient operation with quantum repeaters, which require deterministic, on-demand photon generation for effective entanglement swapping over long distances. This lack of control also affects overall system efficiency, often requiring complex multiplexing schemes to partially mitigate the inherent randomness \cite{Joshi2018Feb, BibEntry2011Apr}.


Epitaxially grown, self-assembled quantum dots (QDs) have shown great potential as sources of on-demand entangled photons, achieving high levels of brightness, purity,  andindistinguishability\cite{liu2019solid,Huber2017May,wang2019demand}. These QDs can be engineered to emit in specific telecom bands, aligning well with the low-loss transmission windows of optical fibers, which is crucial for quantum communication applications over long distances\cite{Schweickert2018Feb,Vajner2024Feb,Zeuner2021Aug,Lettner2021Dec}. However, despite these advantages, self-assembled quantum dots using lattice-mismatched semiconductors are randomly distributed, which complicates their localization and integration into scalable photonic devices. In contrast, site-controlled nanowire-based QDs present additional advantages for practical applications \cite{Dalacu2012Nov, Versteegh2014Oct, Laferriere2022Apr,gao2024demand}. The nanowire structure not only enables precise spatial control of the quantum dot's position but also serves as a waveguide, enhancing the quantum dot's brightness by efficiently directing emitted photons into the optical mode of the nanowire. This increased brightness improves collection efficiency into optical fibers \cite{Bulgarini2014Jul}, making site-controlled, nanowire-based QDs an attractive option for practical quantum communication systems operating in the telecom O-band. In addition, site-controlled nanowire QDs offer a viable route for scalable integration of quantum dot emitters in advance on-chip photonic structures \cite{Mnaymneh2020Feb,Yeung2023Nov,elshaari2020hybrid,descamps2024acoustic,gao2023scalable,Descamps2023,chang2023nanowire,elshaari2018strain,elshaari2017chip,Zadeh2016}.

In this work, we demonstrate the on-demand generation of entangled photon pairs from site-controlled nanowire-based QDs in the telecom O-band. We perform a comprehensive spectroscopic characterization of the nanowire quantum dot and identify a biexciton-exciton (XX-X) cascade. We perform quantum state tomography to characterize the entangled state generated by the QD and calculate the fidelity and concurrence from the reconstructed density matrix. Our findings underscore the potential of nanowire-based QDs as bright, deterministic sources of telecom entangled photons for scalable and efficient quantum applications.

The quality of the entangled photon pairs generated by the XX-X cascade is strongly influenced by the fine-structure splitting (FSS) of the quantum dot. This splitting arises from asymmetries in the quantum dot’s confinement potential. Such asymmetries can be caused by shape anisotropy or strain-induced distortions, which break the degeneracy of the bright exciton states. As a result, the exciton splits into two linearly polarized components with slightly different energies, leading to a time-dependent phase evolution in the polarization-entangled photon pair. The time-evolved entangled state emitted by a quantum dot with a FSS of $\Delta E_{FSS}$ can be written as:

\begin{equation}
\ket{\psi(t)} = \frac{1}{\sqrt{2}} \big( \ket{H_{\text{XX}} H_{\text{X}}} + e^{i \Delta E_{FSS} t / \hbar} \ket{V_{\text{XX}} V_{\text{X}}} \big),
\end{equation}
where $\hbar$ is the reduced Planck constant.

With fast detection systems, such as superconducting nanowire single-photon detectors (SNSPDs) used in the present work, it is possible to resolve the time-evolved entangled state introduced by the FSS. Although the emitted entangled state evolves over time due to FSS, its fidelity does not decay; rather, the photons remain entangled, albeit in a time-dependent state. Thus, by utilizing a time-resolved scheme all emitted photons can be utilized for certain quantum applications such as quantum key distribution (QKD), where the time evolution of the state is exploited rather than avoided \cite{Pennacchietti2024Feb}. 

In addition, techniques to actively mitigate the effects of FSS further expand the usability of quantum dot-based sources. Strain tuning, for instance, can be employed to symmetrize the quantum dot’s confinement potential, effectively erasing the FSS \cite{Lettner2021Dec}. Alternatively, post-emission processing schemes can be used to correct the state after the photons are emitted, ensuring that all emitted entangled photon pairs are suitable for high-fidelity quantum applications \cite{Fognini2018Sep, Varo2022Mar}.

\section{Method}

\begin{figure*}[htb]
\centering

\includegraphics[width=1\columnwidth]{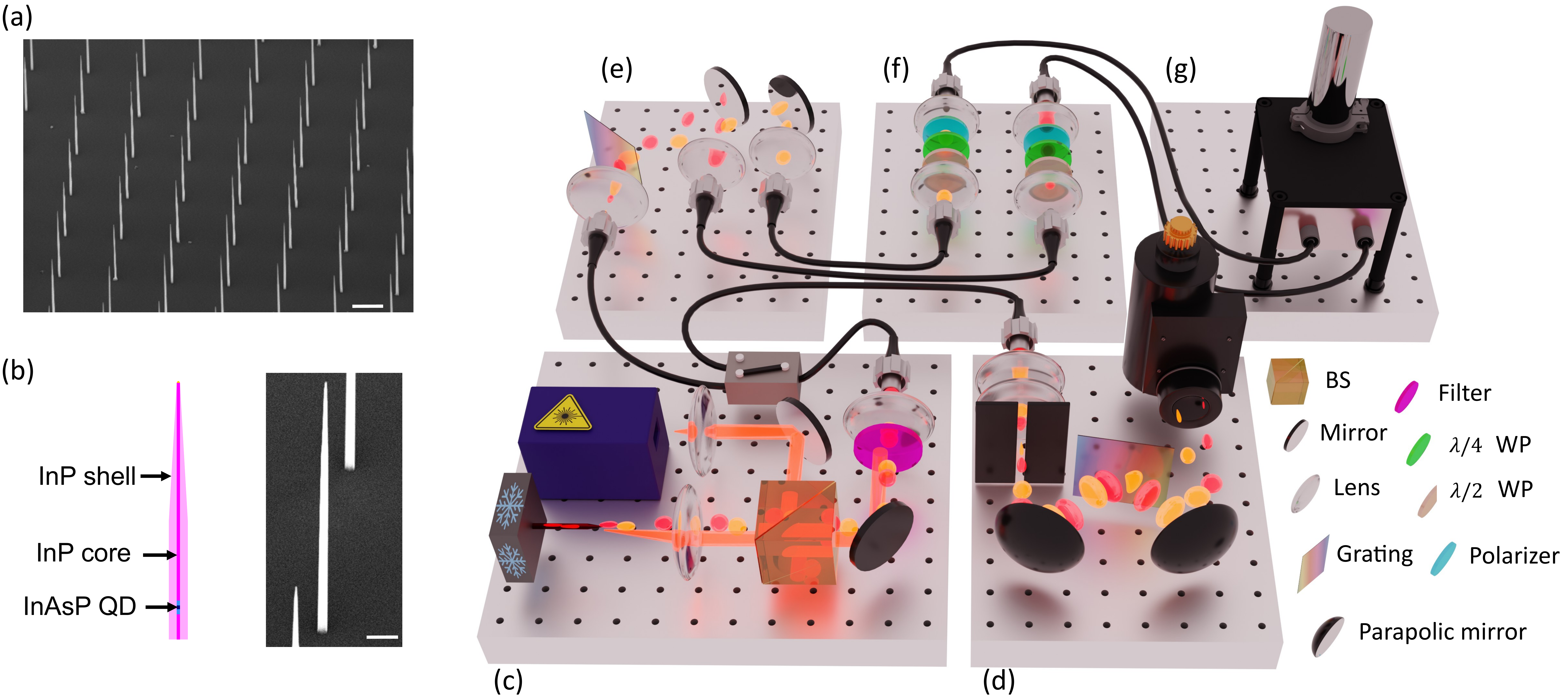}
\caption{\label{fig:fig1-setup}
\textbf{SEM image of the nanowires and schematic of the experimental setup used for QD characterization and quantum state tomography.} (a) SEM image of part of the nanowires array with (b) a higher magnification image of a single nanowire and a schematic showing its structure. Scale bars are 4$\mu$m and 1 $\mu$m, respectively (c) QD excitation by a confocal setup, a power stabilized 80 MHz laser is used to excite the QD into the p-shell. The emission is collected by an objective with NA of 0.8 and coupled into a single-mode fiber. (d) An InGaAs spectrometer for spectral characterization of telecom emission. (e) Transmission grating for filtering the exciton and biexciton and coupling into single mode fibers. (f) Quantum state tomography setup. (g) superconducting nanowire single-photon detectors (SNSPDs).
}
\end{figure*}

The InAsP/InP nanowire quantum dots are grown from the bottom up via chemical beam epitaxy using a selective-area vapor-liquid-solid (SA-VLS) growth technique~\cite{Dalacu_APL2011}. First, an InP core with a diameter of $\sim20$\,nm is grown. The process is briefly interrupted to grow a dot-in-a-rod structure consisting of a single InAs$_{0.68}$P$_{0.32}$ quantum dot of thickness $\sim3$\,nm grown within an InAs$_{0.5}$P$_{0.5}$ nanowire rod of thickness $\sim20$\,nm. Second, the core is cladded with an InP shell to produce a photonic waveguide of base diameter 310\,nm tapered to 20\,nm over the 12\,$\mu$m length of the waveguide~\cite{Laferriere_NL2023}. A scanning electron microscopy (SEM) image of a section of the nanowire array is shown in Figure \ref{fig:fig1-setup}(a), highlighting the uniformity and spatial arrangement of the nanowires. A higher magnification SEM image of a single nanowire is shown in Figure \ref{fig:fig1-setup}(a), revealing its tapered geometry. The schematic alongside the zoomed-in image provides a conceptual representation of the nanowire structure

The experimental setup, illustrated in Figure \ref{fig:fig1-setup}, is designed to excite the quantum dots and collect the emission through a confocal microscopy setup, Figure \ref{fig:fig1-setup}(c), before directing the collected emission to different characterization modules for further characterization. An 80 MHz picosecond pulsed laser source passes through a 10:90 beam splitter (BS). 10 $\%$ of the excitation light is directed toward the QD sample, while the remaining 90$\%$ is directed towards a photo diode used to feedback to an electronic variable optical attenuator (VOA) for power stabilization. The QDs are housed within an attodry 2100 cryostat operating at a temperature of 1.6 K. The sample scanning is achieved using cryogenic nano-positioners. An objective lens with 0.8 NA focuses the laser onto the QD and collects the emitted photons.

The emitted light from the QD passes through a 1250 nm long-pass filter to reject the excitation light and ensure only the QD emission is collected into a single mode SMF-28 fiber. The filtered light is then directed either to a spectrometer equipped with a liquid nitrogen-cooled InGaAs diode detector for photoluminescence measurements, Figure \ref{fig:fig1-setup}(d), or to a transmission Grating (TG) setup, Figure \ref{fig:fig1-setup}(e), to spectrally filter the X and XX and couple them into separated single mode fibers.

The auto-correlation measurements were performed using a 50:50 fiber splitter to split the spectrally filtered emission into two paths, with each path directed to an SNSPD. The detected events from both SNSPDs are then time-stamped and digitized using a quTAG time-to-digital converter (TDC), enabling precise measurement of the arrival time correlations between the photons. 

For entanglement measurements, the filtered XX and X photons are each directed into a polarization analysis setup consisting of a quarter-wave plate, a half-wave plate, and a polarizer, Figure \ref{fig:fig1-setup}(f). The wave plates are mounted on motorized rotational stages, enabling precise control over the polarization basis. A full quantum state tomography is performed by measuring a total of 36 projection bases to reconstruct the density matrix and accurately evaluate the entanglement fidelity and concurrence of the entangled photon pairs.

\section{Results}

\begin{figure*}[htb]
\centering

\includegraphics[width=\columnwidth]{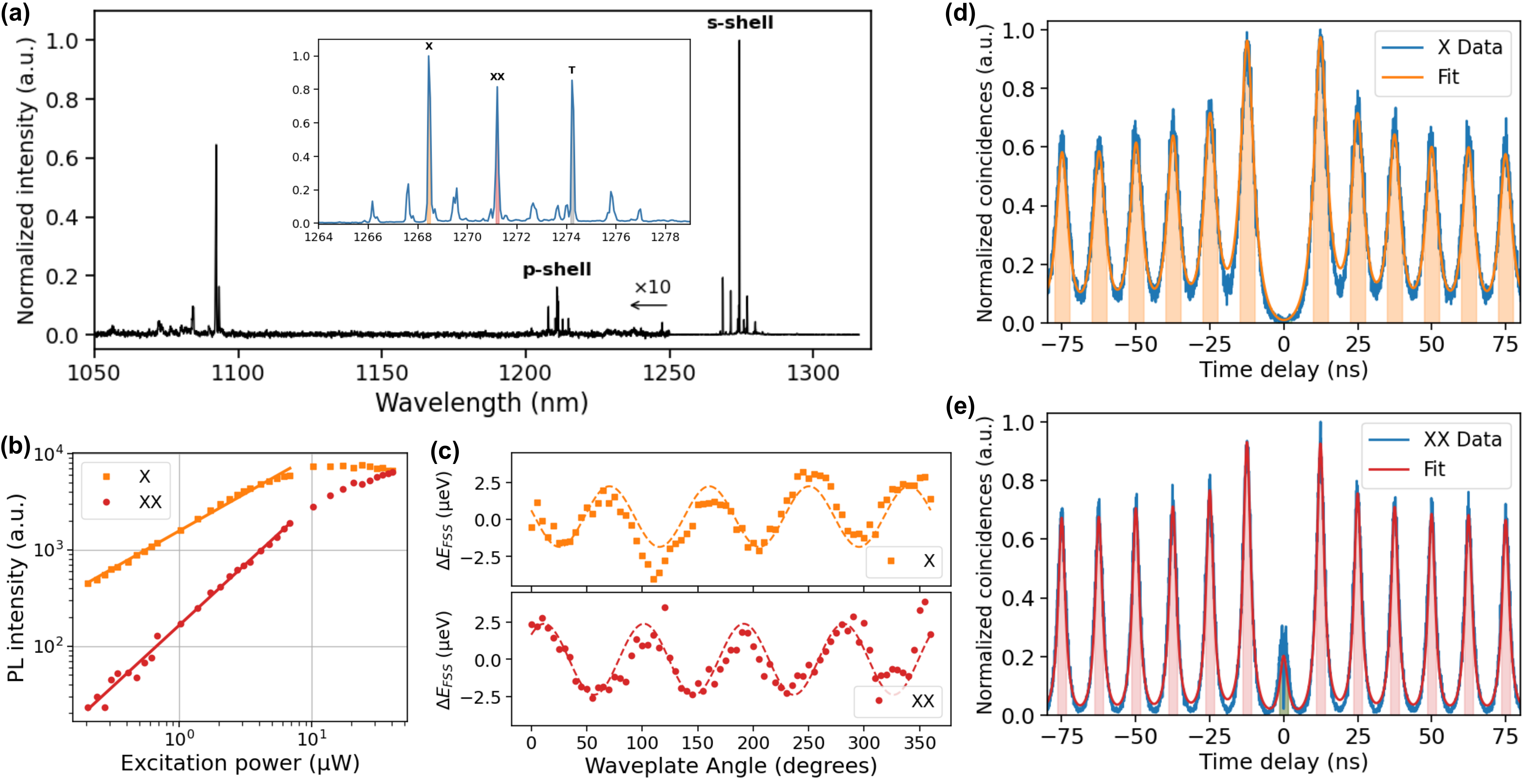}
\caption{\label{fig:fig2-spectra}
\textbf{Spectroscopic characterization of the quantum dot emission.} (a) The QD emission spectrum measured with an InGaAs spectrometer when exciting above band at 793 nm. The insert shows the three most prominent peaks, when exciting into the p-shell, corresponding to the exciton X (orange), biexciton XX (red) and trion T (gray). (b) Power dependence measurement of the X and XX peaks. The exciton shows a sub-linear dependence before it saturates around 10 $\mu W$, while for the biexciton, we observe a close to linear dependence. (c) Polarization-dependent measurement for both X (orange) and XX (red) to determine the fine-structure splitting. The average splitting extracted from the sinusoidal fit (dashed line) is 4.6 $\mu eV$. (d) and (e) are measurements of the autocorrelation function $g^2(\tau)$ for X and XX, respectively. The solid line is the summed Lorentzian fit of the peaks, with the shaded region indicating the full width at half maximum of each peak. 
}
\end{figure*}

We first measured the photoluminescence properties of the nanowire quantum dot to characterize the excitonic states under above-band excitation at 793 nm wavelength. The PL spectrum is shown in Figure \ref{fig:fig2-spectra}(a). From the spectrum, we can resolve emission from the s-shell and the p-shell. The inset shows the QD emission when excited into the p-shell at 1190 nm, with the three highlighted peaks corresponding to the exciton (X), biexciton (XX), and charged exciton (T) states. These peaks are spectrally well-resolved, allowing for clear identification and spectral filtering of each emission line for further characterization. It should be noted that the wavelength used for p-shell excitation is slightly blue shifted, as compared to the p-shell emission observed around 1210 nm when exciting below the bandgap as depicted in Figure \ref{fig:fig2-spectra}(a). When the excitation was tuned to 1210 nm, no emission was observed. This suggests that direct excitation to the p-shell around 1190 nm wavelength is inefficient, likely due to weak dipole coupling or symmetry-related selection rules. Instead, the blue-shifted excitation at 1190 nm efficiently couples to higher-energy states within the p-shell manifold, which then relax non-radiatively to populate the lower-energy p-shell states, producing the observed emission \cite{Suzuki2018Nov}. This observation underscores the significant role of phonon interactions in the excitation and relaxation dynamics of p-shell excitons which should be investigated further.

Figure \ref{fig:fig2-spectra}(b) shows the power dependence of these peaks, where the maximum intensity of X and XX lines are plotted against the excitation power. The exciton exhibits a sublinear dependence with a slope of 0.78, and saturates at an $P_{exc} =$ 10 $\mu$W. The biexciton, on the other hand, displays a closer-to-linear dependence with a slope of 1.27. Both observed dependencies are below what is often reported in literature for X, which should be linear and for XX, which should be quadratic \cite{10.1063/5.0045880}. The reason why we observed these power relations could be explained by the existence of other competitive radiative decay paths formed by the charged exciton and charged biexciton. A more detailed explanation of this will be presented in the discussion section.

We characterized the FSS of the quantum dot by placing a motorized waveplate followed by a polarizing beam splitter in front of the spectrometer. The waveplate angle was rotated in a full cycle with steps of 5 degrees, recording the PL spectrum at each step. As the waveplate rotated, the polarization component of the X and the XX emission passing through the polarizing beam splitter varies. This allows us to resolve the polarization-dependent energy shifts of the X and XX peaks, revealing the FSS. The observed splitting, indicative of asymmetries in the quantum dot confinement potential, is a critical parameter that influences the entanglement quality of emitted photon pairs. Figure \ref{fig:fig2-spectra}(c) shows the measured FSS oscillation of the X peak (orange) and XX peak (red) as the waveplate rotates. The oscillation was fitted to a sinusoidal function (dashed lines), and the FSS extracted from the fitting was 4.6 $\mu$eV.

To assess the purity of exciton and biexciton, we measured the auto-correlation function $g^2(\tau)$ as shown in Figure \ref{fig:fig2-spectra}(c) and (e) for X and XX, respectively. This was performed with pulsed excitation to the p-shell at 1190 nm. The zero-time-delay autocorrelation function, $g^2(0)$, was determined by dividing the coincidence counts within a time window $\Delta$, centered at zero delay, by the average of the summed coincidence counts within the same time window centered at each of the side peaks. The width of the time window, $\Delta$, was obtained by fitting each side peak to a Lorentzian function and using the average full width at half maximum (FWHM) from these fits. The obtained $g^2(0)$ value for X was 0.024, indicating high single-photon purity. In contrast, the biexciton (XX) exhibited lower purity with a $g^2(0)$ value of 0.38. This higher $g^2(0)$ for XX likely arises from a recapturing process under saturation excitation, evident by the narrow antibunching dip at zero delay. A more detailed discussion of this process can be found in the section S3 in the supplementary material.

\begin{figure*}[htb]
\centering

\includegraphics[width=\columnwidth]{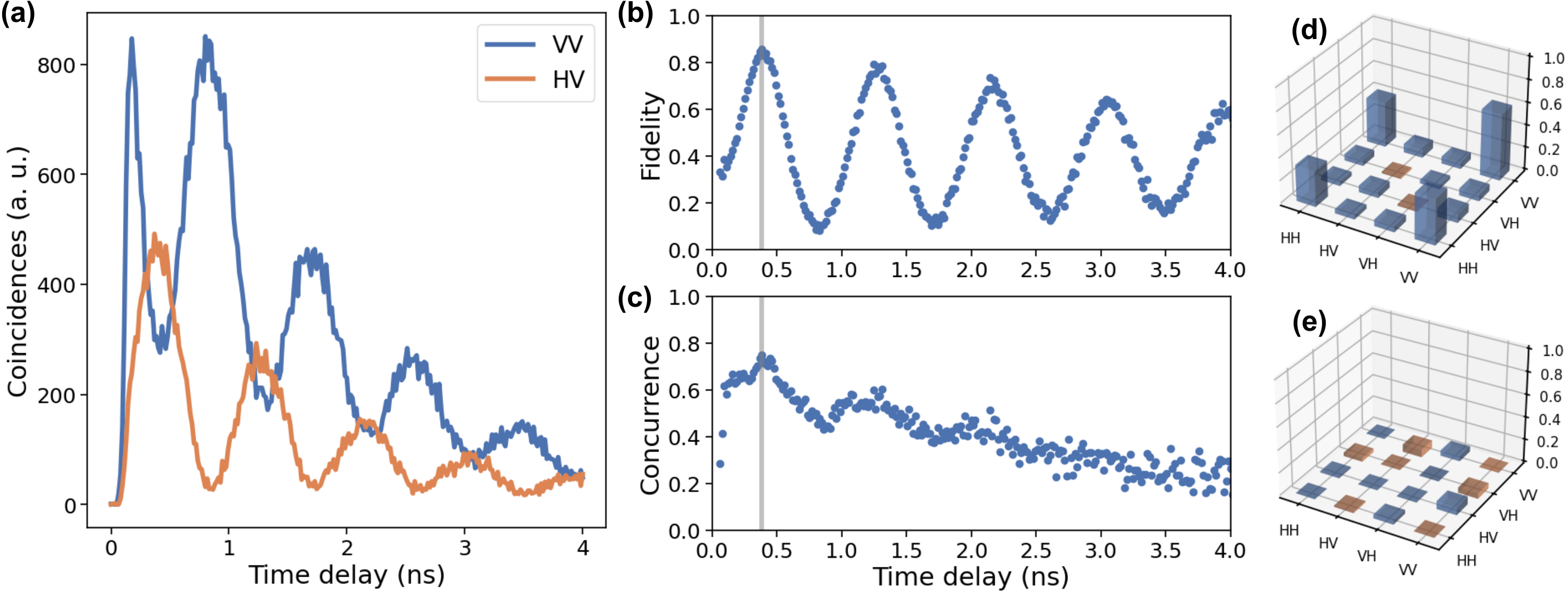}
\caption{\label{fig:fig3-ent}
{\textbf{Quantum state tomography of the telecom O-band entangled photons.} (a) Coincidence measurement of the VV and HV basis. (b) The time evolved fidelity to the $\Phi^+$ state. The oscillation is due to the fine structure splitting of the quantum dot. A maximum fidelity of $85.8\% \pm 1.1\%$ is reached at the time bin highlighted in gray. (c) The reconstructed concurrence at different time bins across the cascade. A maximum concurrence of $75.1\% \pm 2.1\%$ is reached at the same highlighted time bin. (d) and (e) are the real and imaginary parts of the reconstructed density matrix at the highlighted time bin.}
}
\end{figure*}

Quantum state tomography was performed by measuring 36 polarization projections. The analysis utilized the maximum likelihood estimation algorithm in Ref. \cite{afognini2025Jan}, which was modified to reconstruct the density matrices at each time bin across the cascade. Before measuring on the quantum dot, a calibration measurement was conducted using a classical V-polarized laser. The reconstructed state achieved a fidelity of 99.8$\%$ to the ideal V state, confirming the accuracy of the tomography setup. Details of this calibration measurement are provided in section S4 in the supplementary material. Figure \ref{fig:fig3-ent}(a) displays the time-resolved cascade for the VV and HV polarization projections, revealing oscillations in the polarization state attributed to the FSS. The oscillation period is 890 ps, corresponding to an FSS of 4.65 $\mu$eV, in close agreement with the value extracted from the polarization-dependent measurement in Figure \ref{fig:fig2-spectra}(c). To account for systematic polarization rotations and birefringence in our setup, a waveplate rotation with $\theta = 0.521$ rad, and $\phi = -1.284$ rad, derived from the code was applied to the raw density matrices at all time bins. The fidelity to the ideal Bell state $\phi^+$, calculated from the corrected matrices, reached a maximum of $85.8\% \pm1.1\%$ at the time bin highlighted in Figure \ref{fig:fig3-ent}(b) and (c), where the concurrence peaked at $75.1 \%\pm2.1 \%$ as shown in Figure \ref{fig:fig3-ent}(c). The real and imaginary components of the reconstructed density matrix at the highlighted time bin are shown in Figure \ref{fig:fig3-ent}(d) and (e), respectively, exhibiting characteristic features of a highly entangled state with the real part prominently featuring the outer diagonal terms, while the imaginary part exhibits negligible contributions.
The measured fidelity could be further improved by utilizing resonant excitation methods, such as two-photon resonant excitation or phonon-assisted excitation, instead of the p-shell excitation used here. Consequently, this would ensure a more coherent population of the biexciton state and reduce exciton-biexciton timing jitter.


\section{Discussion}



In the investigated nanowire system, different excitation schemes result in different power dependencies. Under p-shell excitation, the power series for X and XX exhibits significantly smaller slopes than expected as was shown in Figure \ref{fig:fig2-spectra}(b). For comparison, we investigated the power dependence of the photoluminescence (PL) under above-bandgap excitation. Instead of displaying the expected quadratic power relation for the biexciton emission, a closer to quadratic dependence is observed for another emission peak, which is attributed to the charged biexciton \cite{Witek2011Nov}. 
This suggests that there are other radiative paths in excitonic system that compete with the primary cascaded emission process. Since our nanowires have a negative background doping, it is more likely to form a charged-biexciton than a neutral biexciton. A power dependence closer to a quadratic relation is more readily observed from the charged biexciton state. Under p-shell excitation, where carriers are predominately generated around the QD, more emission peaks are observed compared to above-bandgap excitation. This further indicates that these alternative radiative pathways significantly influence the carrier dynamic in the nanowire system. A more detailed discussion of this measurement can be found in the supplementary material, sections S1 and S2.


Another noteworthy result is the strong recapturing effect observed in the auto-correlation measurement of the biexciton, which is absent in the exciton. This indicates that in a carrier-rich environment, under saturated excitation power, the QD tends to recapture another electron-hole pair to form the biexciton state after emitting a biexciton photon \cite{Wakileh2024Jan}. By fitting the central anti-bunching dip with a single exponential decay, the recapturing time constant for forming a biexciton state was determined to be approximately 546 ps (see supplementary material section S3). 

\section{Conclusion}

To conclude, we have demonstrated the on-demand generation of polarization entangled photon pairs in the telecom O-band using a nanowire quantum dot. To the best of our knowledge, this represents the first measurement of telecom wavelength entanglement from nanowire quantum dots. The high brightness and low FSS of the dot allowed us to explore and characterize the entanglement properties of the photon pairs.
By performing quantum state tomography, we obtained a maximum fidelity of 85.8$\%$ and a concurrence of 75.1$\%$, demonstrating a high degree of entanglement in the biexciton-exciton cascaded emission. These results were achieved using p-shell excitation, and even higher levels of entanglement could potentially be reached by incorporating resonant excitation techniques.
These results highlight the potential of nanowire quantum dots as bright, deterministic sources of entangled photons for scalable quantum communication and computing applications at telecom wavelengths, providing a foundation for their integration into advanced photonic devices and long-distance quantum networks. 
\section{Acknowledgments}
This research was supported by  Sweden's Innovation Agency VINNOVA (Grant No. 2024-00515), and by the National Sciences and Engineering Research Council of Canada (NSERC).

\bibliography{met_sync}
\end{document}


\newpage

\section{S1. Spectral characterization of the quantum dot}
\label{SM1}

\begin{figure*}[htb]
\centering

\includegraphics[width=0.7\columnwidth]{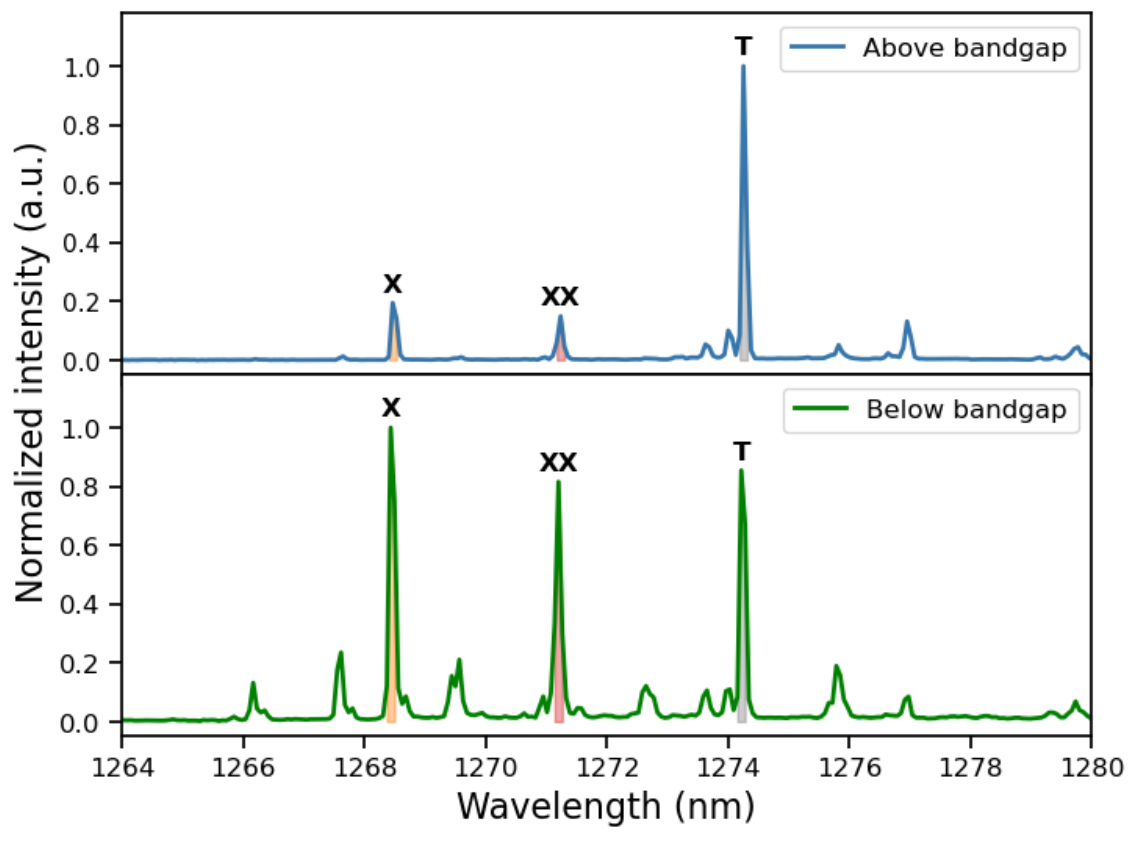}
\caption{\label{fig:SM_fig-spectra}
\textbf{Quantum dot emission at different excitation wavelengths.} The upper plot is for aboveband exciation at 793 nm. The lower plot is for belowband exciation at 881 nm wavelength}
\end{figure*}

Here, we provide a detailed examination of the nanowire quantum dot emission under varying excitation conditions.
Figure \ref{fig:SM_fig-spectra} illustrates the emission spectrum when exciting the quantum dot above the bandgap at 793 nm, and below the bandgap at 881 nm. We observe a strong suppression of the charged exciton state and a significant enhancement of the neutral exciton states when exciting below the band as compared to aboveband.

\begin{figure*}[htb]
\centering

\includegraphics[width=0.7\columnwidth]{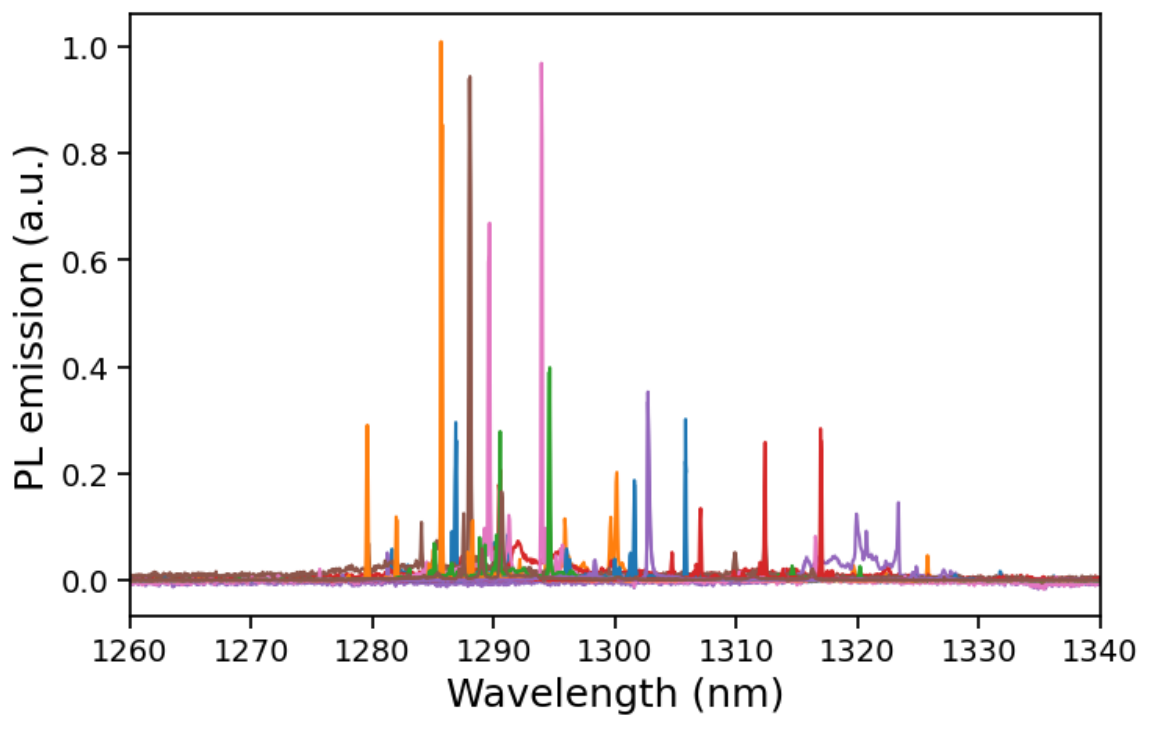}
\caption{\label{fig:SM_fig-Pl_emissions}
\textbf{Photoluminescence spectra of 12 randomly selected dots from the same sample under aboveband excitation.} Different nanowires are represented in different colors.}
\end{figure*}

This difference arises from the distinct carrier generation and capture mechanisms in aboveband and belowband excitation. Aboveband excitation generates free carriers in the host material, which then relax into the quantum dot through scattering or phonon-assisted processes. This often results in an imbalance between electron and hole capture rates, increasing the likelihood of forming charged exciton states due to excess free carriers. In contrast, belowband excitation at 881 nm directly injects carriers into localized states near the quantum dot, bypassing the host material and minimizing the generation of excess free carriers. A similar observation is made with p-shell excitation, where carriers are excited directly into higher quantum dot energy levels before relaxing into the ground state. Both belowband and p-shell excitation provide controlled carrier injection, which suppresses the formation of charged excitons and enhances the emission of neutral exciton states. These observations confirm that excitation schemes targeting quantum dot states more directly minimize the generation of excess free carriers in the host material, suppressing unintended recombination processes and promoting the selective emission of neutral exciton states.

To evaluate the spectral properties and uniformity of the nanowire quantum dots, we measured the photoluminescence spectra of 12 randomly selected dots from the same sample when excited above the band. As shown in Figure \ref{fig:SM_fig-Pl_emissions}, each spectrum, represented by a different color, corresponds to an individual nanowire quantum dot. All measured quantum dots, exhibit emission within the telecom O-band, indicating a high degree of spectral consistency across the sample. This confirms that the quantum dot investigated in the main text is representative of the broader distribution of emitters. 


\begin{figure*}[htb]
\centering

\includegraphics[width=0.7\columnwidth]{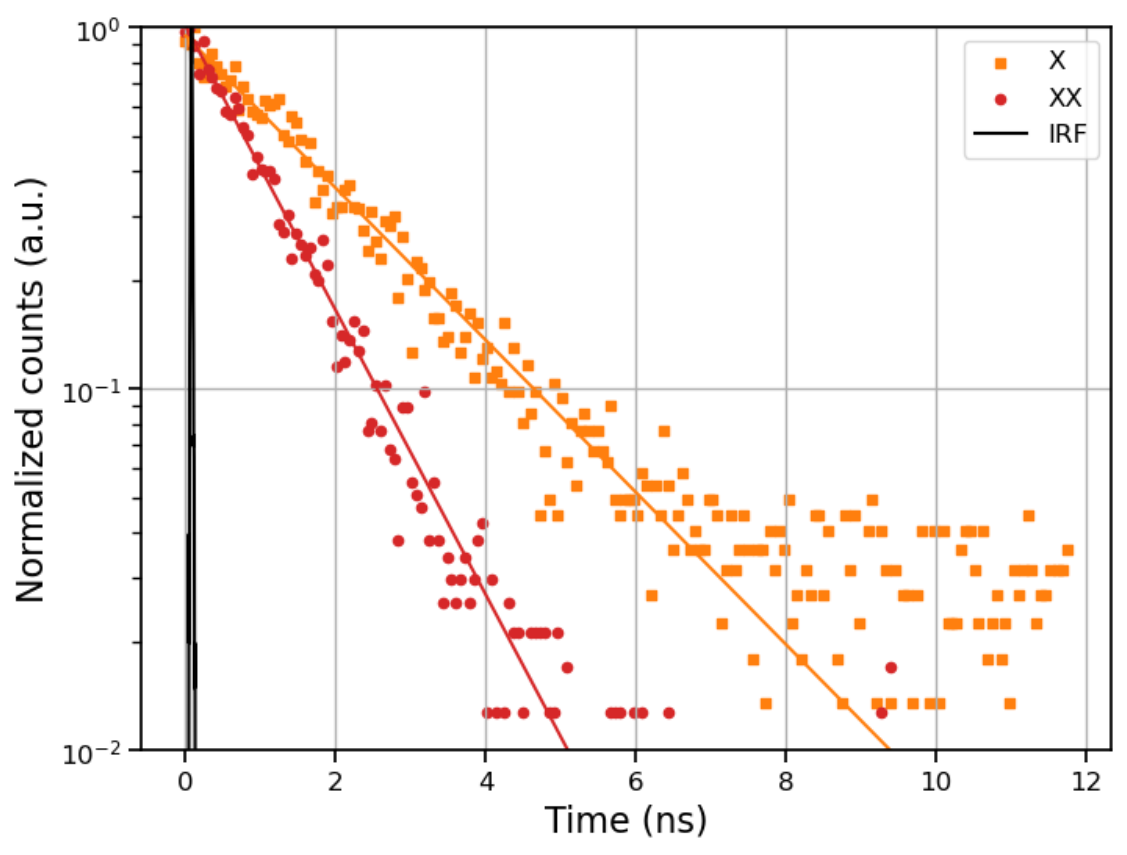}
\caption{\label{fig:SM_fig-SM_lifetime}
\textbf{Lifetime of X and XX under p-shell excitation} The lifetime of X and XX, extracted from  single exponential fit is 2.06 ns and 1.1 ns, respectively}
\end{figure*}

The radiative lifetime of the X and XX when excited to the p-shell is shown in Figure \ref{fig:SM_fig-SM_lifetime}, as well as the instrument response function (IRF). By fitting the data to a single exponential, we obtain a lifetime of 2.06 ns for X and 1.1 ns for XX. It should be noted that the lifetime extracted from the fitting for the exciton does not directly refer to radiative lifetime of the exciton itself. Because of the cascaded emission process, an exciton photon always comes after biexciton emission. What was measured is actually the time constant of the whole cascaded process. The actual radiative lifetime of the exciton can be extracted from the cross-correlation measurements, which revealed lifetime of 1.61 ns here.

\section{S2. Power dependence of neutral and charged excitonic states}
\begin{figure*}[htb]
\centering

\includegraphics[width=\columnwidth]{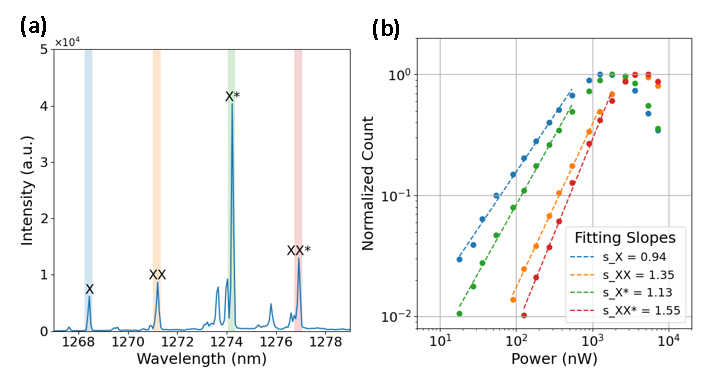}
\caption{\label{fig:SM_fig-SM_NWpowerseries}
\textbf{Power dependence of the investigated nanowire using aboveband excitation} Figure (a) shows the PL spectrum of the nanowire under aboveband excitation. Figure (b) shows the power series corresponding to excitonic emissions.}
\end{figure*}

In the main text, we indicated that other radiative paths in nanowires compete with the conventional cascaded emission, which consequently affects the power dependence of the cascaded photon pairs. Here we further show the power series of the studied nanowire under aboveband excitation. A linear power relation is observed for the exciton and charged exciton, while nonlinear relations are obtained for both biexciton and charged-biexciton cases. We conclude that in the investigated nanowire system, carriers tend to form and radiate from charged-excitonic states, leading to higher PL intensity from the emission of charged excitons. As discussed in Section S1, changing the excitation scheme affects carrier dynamics in the nanowire, ultimately leading to a different power relation under p-shell excitation.


\section{S3. Single-photon purity of Biexciton and recapturing mechanism}
\begin{figure*}[htb]
\centering

\includegraphics[width=0.7\columnwidth]{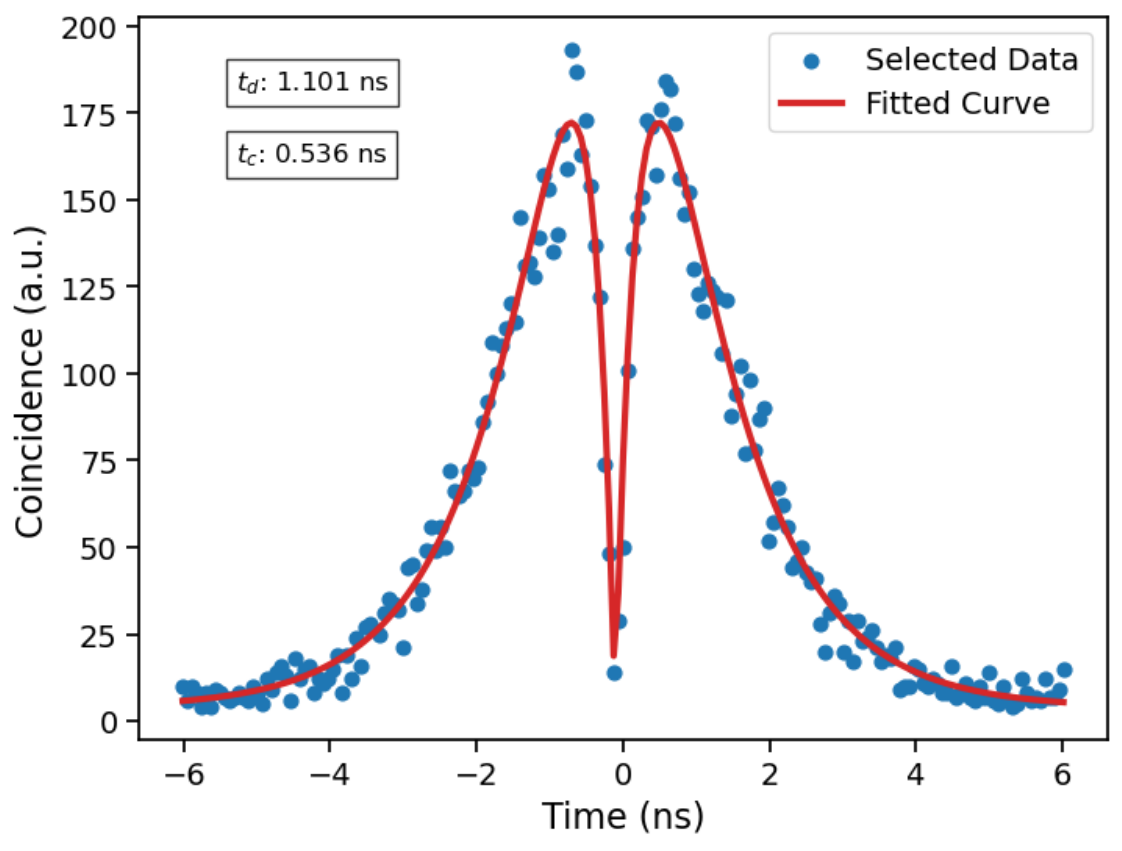}
\caption{\label{fig:SM_fig-XXg2_zoom}
\textbf{Auto-correlation measurement of the biexciton.} A zoomed-in view of the bunching peak at zero time delay. The red line is a fit to the model described in the text.}
\end{figure*}
As shown in the main text, the pulsed $g{^2}(0)$ value calculated from raw data is not as good as expected for the biexciton case. When plotting the data with higher time resolution, there is a clear antibunching dip at zero delay. To explain this phenomenon in detail, we fit the data with the following equation:

\begin{equation}
y(t) = C + A \space \operatorname{exp}\left(\frac{|t-t_0|}{t_d}\right) \left(1-\operatorname{exp}\left(\frac{|t-t_0|}{t_c}\right)\right),
\end{equation}

where $C$ represents the background level contributed by nearby peaks, and $A$ is the amplitude of a double-sided exponential decay peak. Here we extract the time constants of the radiative process $t_d$, and the recapturing process, $t_c$. Due to the nature of the single-photon emitter, the QD should emit only one biexciton photon followed by one exciton photon for each laser pulse, resulting in no coincidence at the zero-delay peak in both the exciton and biexciton cases. However, when we excite the QD at saturation power and provide enough carriers, the QD tends to capture another electron-hole pair to form a biexciton state again, rather than emitting an exciton photon. As a consequence, we observe such recapturing behavior in the biexciton case while maintaining high single-photon purity in the exciton case. It is noteworthy that this effect does not cause any issue for quantum state tomography because each measured photon pair is guaranteed to originate from the same cascaded emission process.


\section{S4. Source brightness and setup efficiency}
The total throughput of the optical setup, including the objective lens, beam splitter, fiber coupling, fiber mating sleeves, and transmission grating, was measured to be 0.8$\%$, with the superconducting nanowire single-photon detector (SNSPD) contributing an additional 50$\%$ efficiency. At an excitation repetition rate of 80 MHz, we measured a photon count rate of approximately 40 kcps for both X and XX when excited into the p-shell. Taking into account the combined efficiency of the optical setup and the SNSPDs, the estimated photon rate from the source at the first lens is approximately 16.67 MHz, which corresponds to 12.5$\%$ of the 80 MHz excitation repetition rate. This demonstrates the high brightness of the nanowire quantum dot source.

\begin{figure*}[htb]
\centering

\includegraphics[width=1\columnwidth]{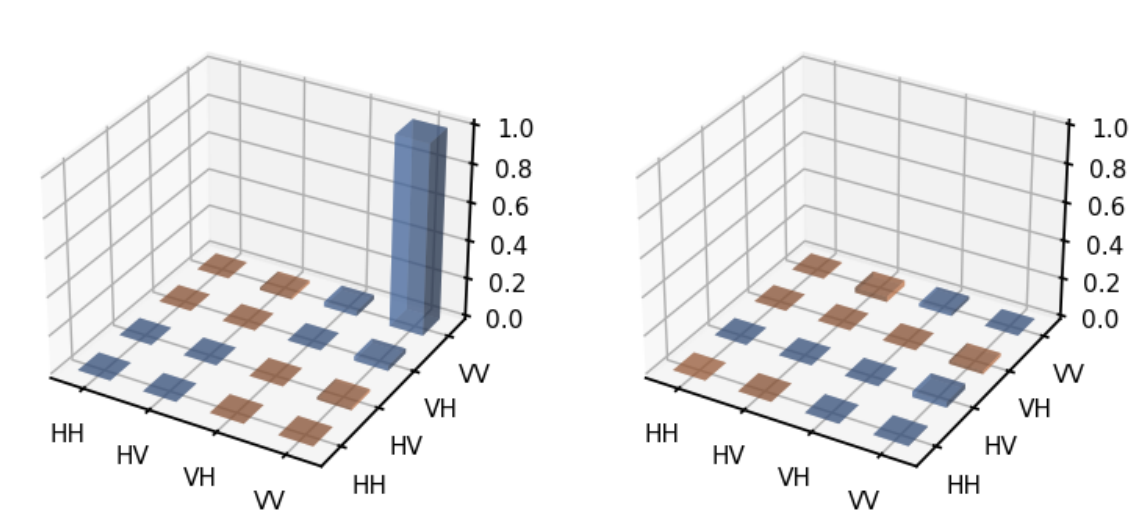}
\caption{\label{fig:SM_fig-dm_VV}
\textbf{A calibration state tomography measured using classical light prepared in a vertical polarization state}}
\end{figure*}

\section{S5. Calibration of the tomography setup}
To validate the accuracy of the quantum state tomography setup used for measuring entanglement generated by a quantum dot, a calibration measurement was performed following the acquisition of projection angles for 36 polarization bases. A laser source, set to a well-defined VV polarization state using a quarter-wave plate and a half-wave plate, was used as the input to the tomography setup. Quantum state tomography was performed with 16 polarization projections, allowing for the reconstruction of the input polarization state based on the code in Ref.\cite{afognini2025Jan}. The reconstructed density matrix from the measured projections is shown in Figure \ref{fig:SM_fig-dm_VV}. We obtain a fidelity of 0.998 ± 0.074 with respect to the ideal VV polarization state, confirming the high accuracy and reliability of the tomography setup for characterizing the entangled state generated by the quantum dot.

\section{S6. Cross-correlation measurements}
Here, we present 16 raw cross-correlation measurements of the XX-X cascade, selected from the 36 measured projections that were used to reconstruct the density matrices at different time bins throughout the cascade. The modulation observed in the cascade is a result of the fine-structure splitting in the quantum dot, which leads to a time-evolved entangled state.
\begin{figure*}[htb]
\centering

\includegraphics[width=1\columnwidth]{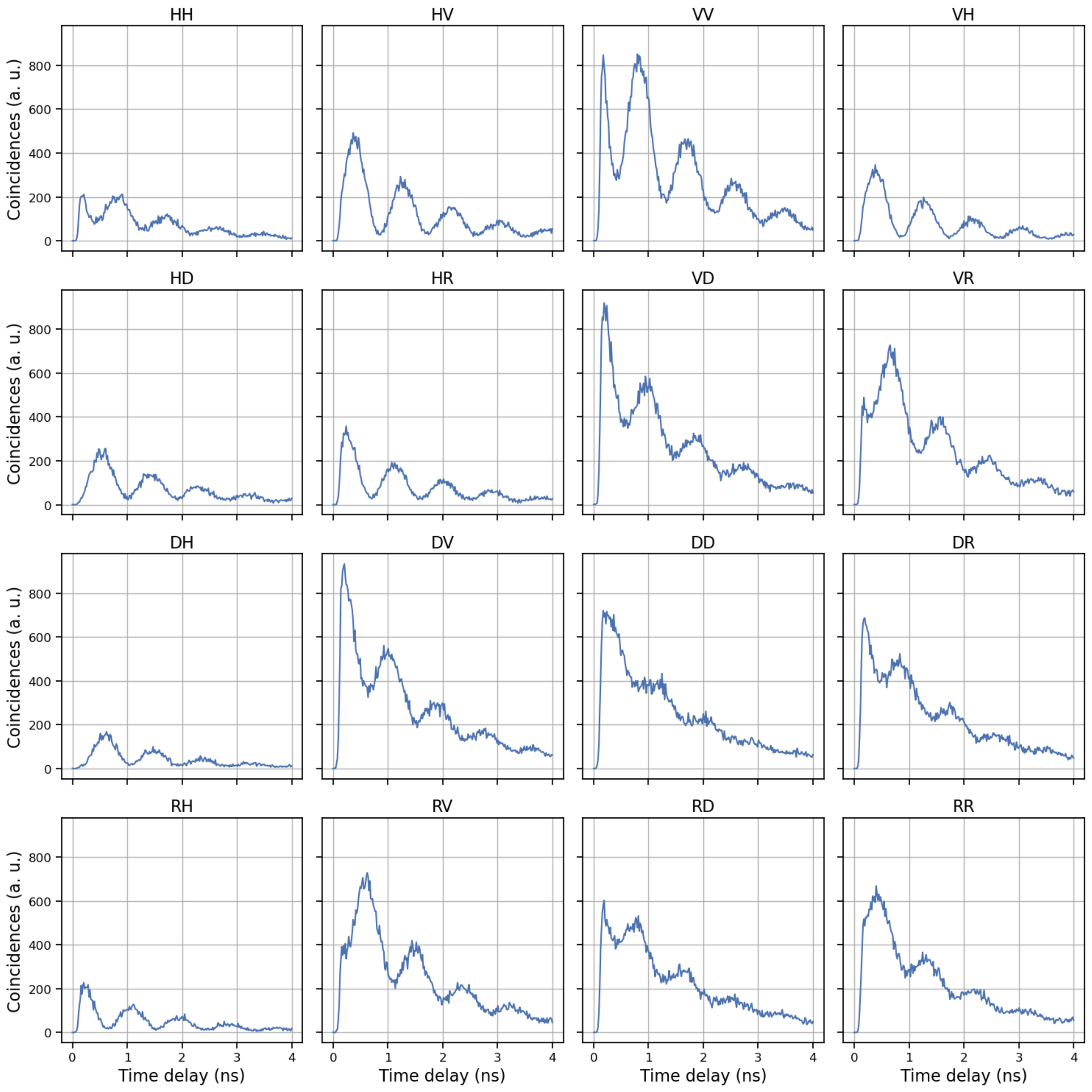}
\caption{\label{fig:SM_fig-Coincidences}
\textbf{Raw cross correlation measurements for 16 polarization projections out of the 36 measured projections.}}
\end{figure*}

\clearpage
\bibliography{met_sync}